\documentclass{ifacconf}

\usepackage{graphicx}      
\usepackage{natbib}        
\usepackage{amsfonts,amssymb}
\usepackage{amsmath}

\newcommand{\R}{\mathbb{R}}

\begin{document}
\begin{frontmatter}

\title{Inversion of Separable Kernel Operators in Coupled Differential-Functional Equations and Application to Controller Synthesis\thanksref{footnoteinfo}}

\thanks[footnoteinfo]{This work was supported by National Natural Science Foundation of PR China under
Grant 61374090, 61503189,  the Natural Science Foundation of Jiangsu Province
under Grant BK20150926. This work was also supported by NSF Grants 1538374, 1301660, 1301851}

\author[First]{Guoying Miao}
\author[Second]{Matthew M. Peet}
\author[Third]{Keqin Gu}

\address[First]{School of Information and Control, Nanjing University of Information Science and Technology,
Nanjing 210044, PR China. (e-mail: mgyss66@163.com).}
\address[Second]{School of Matter, Transport and Energy, Arizona State University, Tempe, AZ 85287, USA. (e-mail: mpeet@asu.edu)}
\address[Third]{Department of Mechanical and Industrial Engineering, Southern Illinois University, Edwardsville, IL 62026, USA. (e-mail: kgu@siue.edu)}

\begin{abstract}                

This article presents the inverse of the kernel operator associated with the complete quadratic Lyapunov-Krasovskii functional for coupled differential-functional equations when the kernel operator is separable. Similar to the case of time-delay systems of retarded type, the inverse operator is instrumental in control synthesis. Unlike the power series expansion approach used in the previous literature, a direct algebraic method is used here. It is shown that the domain of definition of the infinitesimal generator is an invariant subspace of the inverse operator if it is an invariant subspace of the kernel operator. The process of control synthesis using the inverse operator is described, and a numerical example is presented using the sum-of-square formulation.
\end{abstract}


\end{frontmatter}

\section{Introduction}
%
%

It is known that an accurate stability analysis using the Lyapunov approach requires a complete quadratic Lyapunov-Krasovskii functional. Such an approach was first implemented in the form of the discretized Lyapunov-Krasovskii functional method in \cite{Gu:97}, and a refined version was presented in \cite{Gu:01}. In this method, the kernel of the Lyapunov-Krasovskii functional is piecewise linear. An alternative approach is the Sum-Of-Squares (SOS) method presented in \cite{Peet:09}. In the SOS method, the kernel is polynomial. In both approaches, the stability problem is reduced to a semi-definite programming problem, or more specifically, a linear matrix inequality problem.

For many practical systems, the number of state variables with delays is very small compared with the total number of state variables. For such systems, a special form of the coupled differential-difference equation formulation, or its generalized counterpart, the coupled differential-functional equation formulation proves to be much more efficient in numerical computation. The differential-difference formulations can also model systems of neutral type. The discretized Lyapunov-Krasovskii functional approach to stability of differential-difference equations is documented in \cite{Gu:09}, \cite{Gu:10} and \cite{Li:12}, and the SOS formulation can be found in \cite{Zhang:11}.

Control synthesis based on complete quadratic Lyapunov-Krasovskii functional stability conditions is still rare. An early example is \cite{Fridman:02}, in which a more limited class of Lyapunov-Krasovskii functional is used, and some parameter constraints are imposed. Recently, a synthesis based on the inverse of kernel operator associated with the Lyapunov-Krasovskii functional for time-delay systems of retarded type in the SOS formulation was developed in \cite{Peet:09b} and \cite{Peet:13}.

This paper extends the method by Peet et al. to coupled differential-functional equations. The inverse operator is derived using direct algebraic approach rather than the series expansion approach. The basic idea of such synthesis is outlined as follows.

Consider the coupled differential functional equations
\begin{eqnarray}
\dot{x}(t)&=&Ax(t)+By(t-r)+\int_{-r}^{0}H(\theta)y(t+\theta)d\theta, \label{fun} \\
y(t)&=&Cx(t)+Dy(t-r), \label{dif}
\end{eqnarray}
where $A\in \mathbb{R}^{n\times n}$, $B\in \mathbb{R}^{n\times m}$, $C\in \mathbb{R}^{m \times n}$, $H(\theta)\in \mathbb{R}^{n\times m}$, $D\in \mathbb{R}^{m \times n}$, and $r>0$ is time delay, $\mathbb{R}^{n}$ and $\mathbb{R}^{m\times n}$ denote the set of real vectors and matrices with $m\times n$ dimensions, respectively. The initial conditions are defined as
\begin{equation*}
    \begin{array}{rcl}
    x(0)&=&\psi\in \mathbb{R}^{n}\\
y_{0}&=&\phi\in \mathcal{PC}(r,m),
    \end{array}
\end{equation*}
where $y_{\tau}$ represents a shift and restriction of $y(t)$ defined by $y_{\tau}(\theta)=y(\tau+\theta)$,$-r\leq \theta \leq 0$, $\mathcal{PC}(r,m)$ represents the set of piecewise continuous functions from $[-r,0]$ to $\mathbb{R}^m$. Let
\begin{equation}\label{dz1}
Z:=\mathbb{R}^{n}\times \mathcal{PC}(r,m).
\end{equation}
The solutions to the system described by (\ref{fun}) and (\ref{dif}) may be represented by a strongly continuous semigroup($C_0$-semigroup) $\mathcal{S}:Z\rightarrow Z$,
\begin{equation}\label{c1}
z(t)=\mathcal{S}(t-\tau)z(\tau).
\end{equation}
System (\ref{fun})-(\ref{dif}) may be written as an abstract differential equation on $Z$,
\begin{equation}\label{z1}
\dot{z}=\mathcal{A}z,
\end{equation}
where $\mathcal{A}$ is the infinitesimal generator of the $C_0$-semigroup $\mathcal{S}$.
Then the stability of the system can be investigated using a quadratic Lyapunov-Krasovskii functional
\begin{equation}
V(z)=<z,\mathcal{P}z>, \label{V}
\end{equation}
where $\mathcal{P}$ is a self-adjoint operator, and $\left<\cdot,\cdot\right>$ represents inner product. The system is stable if $V(z)$ is positive definite in some sense, and its derivative along the system trajectory
\begin{equation*}
\dot{V}(z)=<z,(\mathcal{PA}+\mathcal{A^*P})z>
\end{equation*}
is negative definite in some sense, where $\mathcal{A^*}$ is the adjoint operator of $\mathcal{A}$. Thus testing the stability of the system can be accomplished by searching for a $\mathcal{P}$ which satisfies the above conditions, and the problem can be reduced to a semi-definite programming problem when the kernel of the integral operator which defines $\mathcal{P}$ is restricted to be either piecewise linear or polynomial because the operator $\mathcal{P}$ appears linearly in $V(z)$ and $\dot{V}(z)$.

However, the situation is quite different for control synthesis. Consider a system with input described by the abstract differential equation
\begin{equation}\label{c2}
\dot{z}=\mathcal{A}z + \mathcal{F}u.
\end{equation}
If we want to design a linear feedback control in the form of
\begin{equation}
u=\mathcal{K}x,
\end{equation}
so that the closed-loop system is stable, and use the Lyapunov-Krasovskii functional given in (\ref{V}), then the derivative becomes
\begin{equation}
\dot{V}(z)=<z,(\mathcal{PA}+\mathcal{A^*P}
+\mathcal{PFK}+(\mathcal{PFK})^*)z>.
\end{equation}
Because we need to determine the feedback gain $\mathcal{K}$ in addition to the operator $\mathcal{P}$, $\dot{V}(z)$ becomes a bilinear function of the parameters. Determining the existence of parameters to make $V(z)$ positive definite and $\dot{V}(z)$ negative definite poses a formidable numerical problem, for which there is not yet any established reliable method to implement.

One solution to this difficulty is to make a variable transformation
\begin{equation}
\hat{z} = \mathcal{P}z, \label{zhat}
\end{equation}
and use new parameters
\begin{eqnarray}
\mathcal{Q}&=& \mathcal{P}^{-1}, \label{opQ} \\
\hat{\mathcal{K}} &=& \mathcal{KP}^{-1} \label{opKhat}
\end{eqnarray}
instead of $\mathcal{P}$ and $\mathcal{K}$ to express $V$ and $\dot{V}$. It can be easily obtained that
\begin{eqnarray}
V(z) &=& <\hat{z},\mathcal{Q}\hat{z}>, \label{Vhat} \\
\dot{V}(z) &=& <\hat{z}, (\mathcal{AQ}+\mathcal{QA}^*
+\mathcal{F}\hat{\mathcal{K}}+\hat{\mathcal{K}}^{*}\mathcal{F}^{*})\hat{z}>, \label{Vdothat}
\end{eqnarray}

which are linear with respect to the new parameters. Once $\mathcal{Q}$ and $\hat{\mathcal{K}}$ have been determined, the original parameters $\mathcal{P}$ and $\mathcal{K}$ may be obtained by solving (\ref{opQ}) and (\ref{opKhat}), at least symbolically.

Critical to implementing the above idea is the inversion of the linear operator $\mathcal{P}$. Unfortunately, such an inversion is not easy in general. It turns out that a relatively simple expression for $\mathcal{P}^{-1}$ is possible when $\mathcal{P}$ is separable, as is utilized in \citep{Peet:09b} to carry out control synthesis for time-delay systems of retarded type. In this paper, we present the inversion of $\mathcal{P}$ for coupled differential-functional equations. Unlike the series expansion method used in \citep{Peet:13}, a direct algebraic approach is used here. Control synthesis is also described, and a numerical example is presented to illustrate the method.

\section{Preliminaries}
Consider the coupled differential functional equations given in (\ref{fun}) and (\ref{dif}).

Stability of such a system may be verified using a complete Lyapunov-Krasovskii functional of the following form,
\begin{eqnarray}\label{34}
V(\psi,\phi)&=&r \psi^{T}P\psi+2r\psi^{T}\int_{-r}^{0}Q(\eta)\phi(\eta)d\eta\nonumber\\
&&+\int_{-r}^{0}\int_{-r}^{0}\phi^{T}(\xi)R(\xi,\eta)\phi(\eta)d\xi d\eta\nonumber\\
&&+\int_{-r}^{0}\phi^{T}(\eta)S(\eta)\phi(\eta)d\eta,
\end{eqnarray}
where
\begin{eqnarray}
P&=&P^{T}\in \mathbb{R}^{n\times n},\label{v1}\\
Q(\eta)&\in& \mathbb{R}^{n\times m},\label{v2}\\
R(\xi,\eta)&=&R^{T}(\eta,\xi)\in \mathbb{R}^{m\times m},\label{v3}\\
S(\eta)&=&S^{T}(\eta)\in \mathbb{S}^{n},\label{v4}
\end{eqnarray}
and $\mathbb{S}^{n}$ represents the set of symmetric matrices, the superscript $T$ denotes the transpose of a matrix or vector.
\begin{lem}\label{lem1}
(Gu and Liu, 2009; Li, 2012) System (\ref{fun})-(\ref{dif})
 with $\rho(D)<1$ is exponentially stable if  there exists a quadratic Lyapunov-Krasovskii functional from (\ref{34})-(\ref{v4}), such that
\begin{equation*}
\epsilon||\psi||^{2}\leq V(\psi,\phi),
\end{equation*}
for some $\epsilon>0$, and its derivative along the system trajectory
\begin{eqnarray*}
&&\dot{V}(\psi,\phi)\nonumber\\
&\triangleq& \limsup\limits_{t\rightarrow 0^{+}}\frac{V(t,x(t,\psi,\phi),y_{t}(\psi,\phi))-V(\psi,\phi)}{t}
\end{eqnarray*}
satisfies
\begin{equation*}
\dot{V}(\psi,\phi)\leq-\epsilon||\psi||^{2},
\end{equation*}
where, $||\psi||$ denotes 2-norm.
\end{lem}

Define inner product on $Z$ given in (\ref{dz1}),
\begin{eqnarray*}
\left\langle\left[\begin{array}{c}\psi_{1}\\
\phi_{1}\end{array}\right],\left[\begin{array}{c}\psi_{2}\\
\phi_{2}\end{array}\right]\right\rangle=r \psi_{1}^{T}\psi_{2}+\int_{-r}^{0}\phi_{1}^{T}(s)\phi_{2}(s)ds.
\end{eqnarray*}

For matrix $P$ and matrix functions $Q, R, S$ that satisfy (\ref{v1})-(\ref{v4}), we define
the linear operator $\mathcal{P}:Z\rightarrow Z$
\begin{equation}\label{2}
\mathcal{P}\left[\begin{array}{c}\psi\\
\phi
\end{array}
\right]
=\left[\begin{array}{c}
P\psi+\int_{-r}^{0}Q(s)\phi(s)ds\\
\Pi
\end{array}
\right],
\end{equation}
where
\begin{equation*}
\Pi=r Q^{T}(s)\psi+\int_{-r}^{0}R(s,\theta)\phi(\theta)d\theta+S(s)\phi(s).
\end{equation*}
Obviously, $\mathcal{P}$ is a bounded and self-adjoint linear operator in view of (\ref{v1})-(\ref{v4}), and the Lyapunov-Krasovskii functional may be expressed as
\begin{equation*}
V(\psi,\phi)=\left\langle\left[\begin{array}{c}\psi\\
\phi\end{array}\right],\mathcal{P}\left[\begin{array}{c}\psi\\
\phi\end{array}\right]\right\rangle.
\end{equation*}

As mentioned in Section 1, system (\ref{fun})-(\ref{dif}) define a strongly $C_{0}$-semigroup $\mathcal{S}:Z\rightarrow Z$ that satisfies (\ref{c1}). System (\ref{fun})-(\ref{dif}) may also be written as an abstract differential equation (\ref{z1}) on $Z$.

Let the domain of definition of $\mathcal{A}$ be $X$. Then,
\begin{equation*}
X:=\left\{\left[\begin{array}{c}\psi\\
\phi\end{array}\right]\in Z\left| \dot{\phi}(s)\in \mathcal{C},  \phi(0)=C\psi+D\phi(-r)\right.\right\},
\end{equation*}
where $\mathcal{C}$ represents the set of continuous functions. It is of interest in some cases to restrict $\mathcal{P}$ so that $X$ is invariant subspace of $\mathcal{P}$,
\begin{equation}\label{p1}
\mathcal{P}X\in X.
\end{equation}
The specific conditions for such a $\mathcal{P}$ to satisfy is given in the following.
\begin{lem}
$\mathcal{P}$ satisfies (\ref{p1}) if and only if the following conditions are satisfied,
\begin{equation}\label{28}
r Q^{T}(0)+S(0)C=CP+rDQ^{T}(-r),
\end{equation}
\begin{equation}\label{29}
R(0,s)=CQ(s)+DR(-r,s),\quad \forall s,
\end{equation}
\begin{equation}\label{30}
DS(-r)=S(0)D.
\end{equation}
\end{lem}
\begin{pf}
Define $h(s)=r Q^{T}(s)\psi+\int_{-r}^{0}R(s,\theta)\phi(\theta)d\theta+S(s)\phi(s)$ and $g=P\psi+\int_{-r}^{0}Q(s)\phi(s)ds$.
Then, $\mathcal{P}X\in X$ is equivalent to
\begin{equation}\label{36}
h(0)=Cg+Dh(-r),
\end{equation}
for arbitrary $\left[\begin{array}{c}\psi\\
\phi\end{array}\right]\in X$, or
\begin{equation}\label{37}
\phi(0)=C\psi+D\phi(-r).
\end{equation}
Using (\ref{37}), we have
\begin{eqnarray}\label{38}
h(0)&=&r Q^{T}(0)\psi+S(0)\phi(0)+\int_{-r}^{0}R(0,\theta)\phi(\theta)d\theta\nonumber\\
&=&r Q^{T}(0)\psi+S(0)C\psi+\int_{-r}^{0}R(0,\theta)\phi(\theta)d\theta\nonumber\\
&&+S(0)D\phi(-r)\nonumber\\
&=&(r Q^{T}(0)+S(0)C)\psi+\int_{-r}^{0}R(0,\theta)\phi(\theta)d\theta\nonumber\\
&&+S(0)D\phi(-r)
\end{eqnarray}
and
\begin{eqnarray}\label{39}
&&C g+Dh(-r)\nonumber\\
&=&C P \psi +\int_{-r}^{0}C Q(s)\phi(s)ds+rDQ^{T}(-r)\psi\nonumber\\
&&+\int_{-r}^{0}DR(-r,s)\phi(s)ds+DS(-r)\phi(-r)\nonumber\\
&=&(CP+rDQ^{T}(-r))\psi+DS(-r)\phi(-r)\nonumber\\
&&+\int_{-r}^{0}(CQ(s)+DR(-r,s))\phi(s)ds
\end{eqnarray}
The right sides of (\ref{38})-(\ref{39}) are equal for arbitrary $\psi$ and $\phi$ if and only if (\ref{28})-(\ref{30}) are satisfied.
\end{pf}

Obviously, the above is generalization of Theorem 3 in \cite{Peet:16}.

\section{Inverse Operator}
In this section, we will present an analytical expression for the inverse of the operator $\mathcal{P}$ when it is separable. Similar to \cite{Peet:13}, such an analytic expression for the inverse operator can be used to expedite the construction of the stabilizing controller in the controller synthesis problem.\\
\textbf{Definition 1:} An operator $\mathcal{P}$ defined in (\ref{2}) is said to be separable if
\begin{equation}\label{40}
R(s,\theta)=Z^{T}(s)\Gamma Z(\theta),
\end{equation}
\begin{equation}\label{41}
Q(s)=HZ(s),
\end{equation}
for some constant matrices $\Gamma=\Gamma^{T}$ and $H$, and column vector function $Z(s)$.
\begin{thm}\label{thm1}
Assume $\mathcal{P}$ in (\ref{2}) is separable.
Then, provided that all the inverse matrices below are well defined, its inverse may be expressed as
\begin{eqnarray}\label{42}
&&\mathcal{P}^{-1}\left[\begin{array}{c}\psi\\
\phi\end{array}\right](s)\nonumber\\
&=&\left[\begin{array}{c}\hat{P}\psi+\int_{-r}^{0}\hat{Q}(\theta)\phi(\theta)d\theta\\
r\hat{Q}^{T}(s)\psi+\hat{S}(s)\phi(s)+\int_{-r}^{0}\hat{R}(s,\theta)\phi(\theta)d\theta
\end{array}\right],
\end{eqnarray}
where
\begin{equation}\label{43}
\hat{R}(s,\theta)=\hat{Z}^{T}(s)\hat{\Gamma}\hat{Z}(\theta),
\end{equation}
\begin{equation}\label{44}
\hat{Q}(\theta)=\hat{H}\hat{Z}(\theta),
\end{equation}
\begin{equation}\label{45}
\hat{S}(s)=S^{-1}(s),
\end{equation}
\begin{equation}\label{46}
\hat{Z}(s)=Z(s)S^{-1}(s),
\end{equation}
\begin{equation}\label{6}
\hat{H}=-P^{-1}H T ,
\end{equation}
\begin{equation}\label{6p5}
T=(I+K\Gamma-r KH^{T}P^{-1}H)^{-1},
\end{equation}
\begin{equation}\label{7}
\begin{array}{rcl}
\hat{P}&=&[I+r P^{-1}H T KH^{T}]P^{-1},
\end{array}
\end{equation}
\begin{equation}\label{8}
\begin{array}{rcl}
\hat{\Gamma}&=&[rT^{T} H^{T}P^{-1}H-\Gamma](I+K\Gamma)^{-1},
\end{array}
\end{equation}
\begin{equation}\label{47}
K=\int_{-r}^{0}Z(s)S^{-1}(s)Z^{T}(s)ds,
\end{equation}
and $I$ denotes the identity matrix with appropriate dimension.
\end{thm}
\begin{pf}
Let the operator defined by the right hand side of (\ref{42}) be denoted as $\hat{\mathcal{P}}$, then
\begin{equation*}\label{9}
\hat{\mathcal{P}}\mathcal{P}\left[\begin{array}{c}\psi\\
\phi\end{array}\right](s)=\left[\begin{array}{c}\Lambda_{1}\\
\Lambda_{2}
\end{array}\right],
\end{equation*}
where
\begin{equation*}
\begin{array}{rcl}
\Lambda_{1}&=&\int_{-r}^{0}\left(\hat{P}Q(\theta)+\hat{Q}(\theta)S(\theta)
+\int_{-r}^{0}\hat{Q}(\xi)R(\xi,\theta)d\xi\right)\nonumber\\
&&\cdot\phi(\theta)d\theta+\left(\hat{P}P+\int_{-r}^{0}r\hat{Q}(\theta)Q^{T}(\theta)d\theta\right)\psi,\nonumber\\
\Lambda_{2}&=&r\left(\hat{Q}^{T}(s)P+\hat{S}(s)Q^{T}(s)+\int_{-r}^{0}\hat{R}(s,\theta)Q^{T}(\theta)d\theta\right)\psi\nonumber\\
&&+\hat{S}(s)S(s)\phi(s)+\int_{-r}^{0}\left(r\hat{Q}^{T}(s)Q(\theta)+\hat{S}(s)R(s,\theta)\right.\nonumber\\
&&\left.+\hat{R}(s,\theta)S(\theta)+\int_{-r}^{0}\hat{R}(s,\xi)R(\xi,\theta)d\xi\right)\phi(\theta)d\theta.
\end{array}
\end{equation*}
Using (\ref{40})-(\ref{41}) and (\ref{43})-(\ref{47}), we obtain
\begin{equation*}
\begin{array}{rcl}
&&\hat{P}P+r\int_{-r}^{0}\hat{Q}(\theta)Q^{T}(\theta)d\theta\nonumber\\
&=& \hat{P}P+r\hat{H}K H^{T}\nonumber\\
&=&[I+r P^{-1}H(I+K\Gamma -r KH^{T}P^{-1}H)^{-1}KH^{T}]\nonumber\\
&&-r P^{-1}HTKH^{T}\nonumber\\
&=& I,
\end{array}
\end{equation*}
\begin{equation*}
\begin{array}{rcl}
&&\hat{P}Q(\theta)+\hat{Q}(\theta)S(\theta)+\int_{-r}^{0}\hat{Q}(\xi)R(\xi,\theta)d\xi\nonumber\\
&=&\left(\hat{P}H+\hat{H}+\hat{H}K\Gamma)Z(\theta)\right.\nonumber\\
&=&\left([I+r P^{-1}HTKH^{T}] P^{-1}H-P^{-1}HT\right.\nonumber\\
&&\left.-P^{-1}HTK\Gamma\right)Z(\theta)\nonumber\\
&=&\left[P^{-1}H+P^{-1}HT(r KH^{T}P^{-1}H-I-K\Gamma)\right]Z(\theta)\nonumber\\
&=&\left(P^{-1}H-P^{-1}H\right)Z(\theta)\nonumber\\
&=&0,
\end{array}
\end{equation*}
\begin{equation*}
\begin{array}{rcl}
&&\hat{Q}^{T}(s)P+\hat{S}(s)Q^{T}(s)+\int_{-r}^{0}\hat{R}(s,\theta)Q^{T}(\theta)d\theta\nonumber\\
&=&\hat{Z}^{T}(s)(\hat{H}^{T}P+H^{T}+\hat{\Gamma}KH^{T})\nonumber\\
&=&\hat{Z}^{T}(s)\left\{-T^TH^{T}P^{-1}P+H^{T}\right.\nonumber\\
&&\left.+[rT^T H^{T}P^{-1}H-\Gamma](I+K\Gamma)^{-1}KH^{T}\right\}\nonumber\\
&=&\hat{Z}^{T}(s)\left\{I-T^T (I-r H^{T}P^{-1}H(I+K\Gamma)^{-1}K) \right.\nonumber\\
&&\left.-\Gamma(I+K\Gamma)^{-1}K\right\}H^{T}\nonumber\\
&=&\hat{Z}^{T}(s)\left\{I-T^T(I-r H^{T}P^{-1}HK(I+\Gamma K)^{-1}) \right.\nonumber\\
&&\left.-\Gamma K(I+\Gamma K)^{-1}\right\}H^{T}\nonumber\\
&=&\hat{Z}^{T}(s)\left\{I-T^T(I+\Gamma K-r H^{T}P^{-1}HK)(I+\Gamma K)^{-1}\right.\nonumber\\
&&\left.-\Gamma K(I+\Gamma K)^{-1}\right\}H^{T}\nonumber\\
&=&\hat{Z}^{T}(s)[I-(I+\Gamma K)^{-1}-\Gamma K(I+\Gamma K)^{-1}]H^{T}\nonumber\\
&=&0,
\end{array}
\end{equation*}
\begin{equation*}
\begin{array}{rcl}
&&r\hat{Q}^{T}(s)Q(\theta)+\hat{S}(s)R(s,\theta)+\hat{R}(s,\theta)S(\theta)\nonumber\\
&&+\int_{-r}^{0}\hat{R}(s,\xi)R(\xi,\theta)d\xi\nonumber\\
&=&\hat{Z}^{T}(s)(r\hat{H}^{T}H+\Gamma+\hat{\Gamma}+\hat{\Gamma}K\Gamma)Z(\theta)\nonumber\\
&=&\hat{Z}^{T}(s)\left\{-rT^T H^{T}P^{-1}H+\Gamma\right.\nonumber\\
&&\left.+[rT^T H^{T}P^{-1}H-\Gamma](I+K\Gamma)^{-1}(I+K\Gamma)\right\}Z(\theta)\nonumber\\
&=&\hat{Z}^{T}(s)\left(-rT^T H^{T}P^{-1}H+\Gamma+rT^T H^{T}P^{-1}H\right.\nonumber\\
&&\left.-\Gamma\right)Z(\theta)\nonumber\\
&=&0.
\end{array}
\end{equation*}
Thus, we have shown
\begin{equation}\label{48}
\begin{array}{rcl}
\hat{\mathcal{P}}\mathcal{P}\left[\begin{array}{c}\psi\\
\phi\end{array}\right]=\left[\begin{array}{c}\psi\\
\phi\end{array}\right],
\end{array}
\end{equation}
for all $\left[\begin{array}{c}\psi\\
\phi\end{array}\right]\in Z$.
Similarly, we can show
\begin{equation}\label{49}
\mathcal{P}\hat{\mathcal{P}}\left[\begin{array}{c}\psi\\
\phi\end{array}\right]=\left[\begin{array}{c}\psi\\
\phi\end{array}\right].
\end{equation}
From (\ref{48}) and (\ref{49}), we conclude that $\hat{\mathcal{P}}=\mathcal{P}^{-1}$.
\end{pf}

\begin{thm}
If the separable operator $\mathcal{P}$ satisfies $\mathcal{P}X\in X$, Then, $\mathcal{P}^{-1}X\in X$ holds.
\end{thm}
\begin{pf}
Let the linear operator $\mathcal{P}$ satisfy $\mathcal{P}X\in X$. By Lemma 2, this is equivalent to (\ref{28})-(\ref{30}), from which, we obtain
\begin{eqnarray}
CP^{-1}&=&rS^{-1}(0)(DZ^{T}(-r)-Z^{T}(0))H^{T}P^{-1}\nonumber\\
&&+S^{-1}(0)C,\label{o1}\\
CH&=&(Z^{T}(0)-DZ^{T}(-r))\Gamma,\label{o2}\\
S^{-1}(0)D&=&DS^{-1}(-r).\label{o3}
\end{eqnarray}
Applying (\ref{o1})-(\ref{o3}) to the operator $\mathcal{P}^{-1}$ defined in (\ref{42}), after tedious calculations, we can obtain the following equation,
\begin{equation*}
\begin{array}{rcl}
&&r\hat{Q}^{T}(0)\psi+\hat{S}(0)\phi(0)+\int_{-r}^{0}\hat{R}(0,\theta)\phi(\theta)d\theta\nonumber\\
&=&C\left(\hat{P}\psi+\int_{-r}^{0}\hat{Q}(\theta)\phi(\theta)d\theta\right)+D\left(r\hat{Q}^{T}(-r)\psi\right.\\
&&+\hat{S}(-r)\phi(-r)+\left.\int_{-r}^{0}\hat{R}(-r,\theta)\phi(\theta)d\theta\right),
\end{array}
\end{equation*}
from which, we conclude that $\mathcal{P}^{-1}X\in X$.
\end{pf}
\section{Controller Synthesis}

In this section, we consider a control system as follows
\begin{eqnarray}
\dot{x}&=&Ax+By(t-r)+Fu(t),\label{a1}\\
y(t)&=&Cx(t)+Dy(t-\tau).\label{a2}
\end{eqnarray}
Define the infinitesimal generator $\mathcal{A}$ as follows.
\begin{equation*}\label{12}
\begin{array}{rcl}
\left(\mathcal{A}\left[\begin{array}{c}x\\
y_{t}\end{array}\right]\right)(s)=\left[\begin{array}{c}Ax+By(t-r)\\
\frac{d}{ds}y_{t}(s)\end{array}\right].
\end{array}
\end{equation*}
Likewise, we define the input operator $\mathcal{F}: \mathbb{R}^{q}\rightarrow X$ as
\begin{equation*}
(\mathcal{F}u)(s):=\left[\begin{array}{c}Fu\\0\end{array}\right].
\end{equation*}
We define the controller synthesis problem as the search for matrices $K_0, K_1$ and matrix-valued function $K_2(s)$ such that the System of Equations~\eqref{a1}-~\eqref{a2} is stable if
\begin{equation}\label{13}
u(t)=\mathcal{K}\left[\begin{array}{c}x\\
y_{t}\end{array}\right],
\end{equation}
where we define $\mathcal{K}: X \rightarrow \R^q$ as
\begin{eqnarray}\label{k1}
&&\left(\mathcal{K}\left[\begin{array}{c}x\nonumber\\
y_{t}\end{array}\right]\right)(s)\nonumber\\
&=&K_{0}x(t)+K_{1}y(t-r)+\int_{-r}^{0}K_{2}(s)y(t+s)ds.
\end{eqnarray}
Before we give the main result of the section, we briefly address SOS methods for enforcing joint positivity of coupled multiplier and integral operators using positive matrices. These methods have been developed in a series of papers, a summary of which can be found in the survey paper~\cite{Peet:14}. Specifically, for matrix-valued functions $M(s)$, $N(s,\theta)$, we say that
\[
\{M,N\}\in \Xi
\]
if $M$ and $N$ satisfy the conditions of Theorem 8 in~\cite{Peet:14}. The constraint $\{M,N\}\in \Xi$ can be cast as an LMI using SOSTOOLS as described in~\cite{Peet:14} and this constraint ensures that the operator $\mathcal{P}$, defined as
\begin{equation*}
\mathcal{P}\left[\begin{array}{c}\psi\\
\phi
\end{array}
\right]
=\left[\begin{array}{c}
M_{11}\psi+\int_{-r}^{0}M_{12}(s)\phi(s)ds\\
r M_{21}^{T}(s)\psi+\int_{-r}^{0}N(s,\theta)\phi(\theta)d\theta+M_{22}(s)\phi(s)
\end{array}
\right]
\end{equation*}
is positive on $X$. Furthermore, we note that $\{M,N\}\in \Xi$ implies that $\mathcal{P}$ is separable and $P=\int M_{11}(s)ds$ and $S=M_{22}$ are invertible. We now state the main result.

\textbf{Proposition 1:} Let $Z:\R \rightarrow \R^{q \times n}$ be an arbitrary continuously differentiable function. Suppose there exist matrices $M_{0}$, $M_{1}$, $P=P^{T}$, matrix-valued functions $M_{2}(s)$, $Q(s)$, $R(s,\theta)$, $S(s)=S^{T}(s)\in \mathbb{S}^{n}$, and scalar $\epsilon>0$ such that (\ref{28})-(\ref{30}) are satisfied and the following conditions hold
\[
\{T,R\}\in \Xi, \qquad \{-U,-V\} \in \Xi,
\]
where
\begin{equation}\label{20}
T(s)=\left[\begin{array}{cc} P & r Q(s)\\
r Q^{T}(s) & S(s)
\end{array}
\right] -\epsilon I,
\end{equation}
\begin{align}\label{4}
U(s)=\left[\begin{array}{ccc}
\Gamma+\epsilon I & B S(-r)+F M_{1}+\frac{1}{r}C^{T}S(0)D &\Upsilon\\
 \ast &\frac{-1}{r}(S(-r)-D^{T}S(0)D) & 0\\
  \ast& \ast&\dot{S}(s)
\end{array}\right],
\end{align}
\begin{equation}\label{5}
V(s,\theta)=\frac{d}{ds}R(s,\theta)+\frac{d}{d\theta}R(s,\theta),
\end{equation}
and $\ast$ denotes the corresponding symmetric part,
\begin{equation*}
\begin{array}{rcl}
\Gamma&=&AP+PA^{T}+r(B Q^{T}(-r)+Q(-r)B^{T})\nonumber\\
&&+\frac{1}{r}C^{T}S(0)C+FM_{0}+M_{0}^{T}F,\nonumber\\
\Upsilon&=&r[\dot{Q}(s)+BR(-r,s)+AQ(s)+F M_{2}(s)].
\end{array}
\end{equation*}
Then System (\ref{a1})-(\ref{a2}) is stabilizable with a controller of the form (\ref{13}). Furthermore, let $\hat P$, $\hat Q$, $\hat R$ and $\hat S$ be as defined in Theorem 3. Then if
\begin{equation*}
u(t)=K_{0}x(t)+K_{1}y(t-r)+\int_{-r}^{0}K_{2}(s)y(t+s)ds,
\end{equation*}
where
\begin{align}
K_{0}&=M_{0}\hat P+rM_{1}\hat Q^{T}(-r)+r\int_{-r}^{0}M_{2}(s) \hat Q^{T}(s)ds,\\
K_{1}&=M_{1}\hat S(-r),\\
K_{2}(s)&=M_{0}\hat Q(s)+M_{1} \hat R(-r,s)+M_{2}(s) \hat S(s)\nonumber\\
&+\int_{-r}^{0}M_{2}(\theta)\hat R(\theta,s)d\theta,
\end{align}
then the System (\ref{a1})-(\ref{a2}) is stable.
\begin{pf}
Define
\begin{equation*}
\mathcal{P}\left[\begin{array}{c}\psi\\
\phi
\end{array}
\right]
=\left[\begin{array}{c}
P\psi+\int_{-r}^{0}Q(s)\phi(s)ds\\
r Q^{T}(s)\psi+\int_{-r}^{0}R(s,\theta)\phi(\theta)d\theta+S(s)\phi(s)
\end{array}
\right].
\end{equation*}
Then as per Lemma 2, $\mathcal{P}=\mathcal{P}^*$, $\mathcal{P}: X\rightarrow X$ and $\mathcal{P}\ge \epsilon I$. Furthermore, $\mathcal{P}$ is bounded and as per Theorem 3, the inverse $\mathcal{P}^{-1}$ is defined as in (\ref{42}) and is likewise bounded and coercive with $\mathcal{P}^{-1}\ge \epsilon' I$. Furthermore, from Theorem 4, $\mathcal{P}^{-1}: X\rightarrow X$ and $\mathcal{P}^{-1}=\mathcal{P}^{-*}$. Now define the Lyapunov functional
\begin{equation*}\label{17}
V=\left\langle\left[\begin{array}{c}\psi\\
\phi(s)
\end{array}\right],\mathcal{P}^{-1}\left[\begin{array}{c}\psi\\
\phi(s)
\end{array}\right]\right\rangle \geq \epsilon'   \left|\left|\begin{array}{c}\psi\\
\phi(s)
\end{array}\right|\right|^{2}
\end{equation*}
for $\left[\begin{array}{c}\psi\\
\phi(s)\end{array}\right]\in X$. Since $\mathcal{P}^{-1}=\mathcal{P}^{-*}$,
\begin{equation*}
\begin{array}{rcl}
&&\left\langle\left[\begin{array}{c}\psi\\
\phi(s)
\end{array}\right],\mathcal{P}^{-1}\mathcal{A}\left[\begin{array}{c}\psi\\
\phi(s)
\end{array}\right]\right\rangle\\
&&+\left\langle\mathcal{A}\left[\begin{array}{c}\psi\\
\phi(s)
\end{array}\right],\mathcal{P}^{-1}\left[\begin{array}{c}\psi\\
\phi(s)
\end{array}\right]\right\rangle\\
&=&\left\langle \mathcal{P}^{-1}\left[\begin{array}{c}\psi\\
\phi(s)
\end{array}\right],\mathcal{A}\mathcal{P}\mathcal{ P}^{-1}\left[\begin{array}{c}\psi\\
\phi(s)
\end{array}\right]\right\rangle\\
&&+\left\langle\mathcal{A}\mathcal{ P}\mathcal{ P}^{-1}\left[\begin{array}{c}\psi\\
\phi(s)
\end{array}\right],\mathcal{P}^{-1}\left[\begin{array}{c}\psi\\
\phi(s)
\end{array}\right]\right\rangle.\\
\end{array}
\end{equation*}

Next, we note that if we define $\mathcal{K}$ as
\begin{eqnarray*}
&&\left(\mathcal{K}\left[\begin{array}{c}x\nonumber\\
y_{t}\end{array}\right]\right)\nonumber\\
&=&K_{0}x(t)+K_{1}y(t-r)+\int_{-r}^{0}K_{2}(s)y(t+s)ds,
\end{eqnarray*}
and $\mathcal{M}$ as
\begin{eqnarray*}
&&\left(\mathcal{M}\left[\begin{array}{c}x\nonumber\\
y_{t}\end{array}\right]\right)\nonumber\\
&=&M_{0}x(t)+M_{1}y(t-r)+\int_{-r}^{0}M_{2}(s)y(t+s)ds,
\end{eqnarray*}
then $\mathcal{K}:=\mathcal{M}\mathcal{P}^{-1}$. We construct the controller
\begin{eqnarray*}
&&\mathcal{M}\mathcal{P}^{-1}\left[\begin{array}{c}x\\
y\end{array}\right]\nonumber\\
&=&M_{0}\left(\hat P x+\int_{-r}^{0} \hat Q(\theta)y(\theta)d\theta\right)\nonumber\\
&&+M_{1}\left(r \hat Q^{T}(-r)x+\hat S(-r)y(-r)+\int_{-r}^{0}\hat R(-r,\theta)y(\theta)d\theta\right)\nonumber\\
&&+\int_{-r}^{0}M_{2}(s)\left(r\hat Q^{T}(s)x+\hat S(s)y(s)\right.\nonumber\\
&&\left.+\int_{-r}^{0}\hat R(s,\theta)y(\theta)d\theta\right)ds\nonumber\\
&=&\left(M_{0}\hat P+rM_{1}\hat Q^{T}(-r)+r\int_{-r}^{0}M_{2}(s) \hat Q^{T}(s)ds\right)x\nonumber\\
&&+M_{1}\hat S(-r)y(-r)+\int_{-r}^{0}\left(M_{0}\hat Q(s)+M_{1}\hat R(-r,s)\right.\nonumber\\
&&\left.+M_{2}(s)\hat S(s)+\int_{-r}^{0}M_{2}(\theta)\hat R(\theta,s)d\theta\right)y(s)ds\nonumber\\
&=&\mathcal{K}\left[\begin{array}{c}x\\
y\end{array}\right].
\end{eqnarray*}

Now we define a new state $\left[\begin{array}{c}\hat{\psi}\\
\hat{\phi}(s)\end{array}\right]=\mathcal{P}^{-1}\left[\begin{array}{c}\psi\\
\phi(s)\end{array}\right]\in X$.  Continuing, if $u=\mathcal{K}\left[\begin{array}{c}x\\
y_{t}\end{array}\right]=\mathcal{K}\mathcal{P}\mathcal{P}^{-1}\left[\begin{array}{c}x\\
y_{t}\end{array}\right]=\mathcal{M}\mathcal{P}^{-1}\left[\begin{array}{c}x\\
y_{t}\end{array}\right]$, then the closed-loop system is stable if $\dot{V}<0$, where
\begin{equation*}
\begin{array}{rcl}
\dot{V}&=&\left\langle \left[\begin{array}{c}\hat{\psi}\\
\hat{\phi}(s)
\end{array}\right],\mathcal{A}\mathcal{P}\left[\begin{array}{c}\hat{\psi}\\
\hat{\phi}(s)
\end{array}\right]\right\rangle\nonumber\\
&&+\left\langle\mathcal{A}\mathcal{P}\left[\begin{array}{c}\hat{\psi}\\
\hat{\phi}(s)
\end{array}\right],\left[\begin{array}{c}\hat{\psi}\\
\hat{\phi}(s)
\end{array}\right]\right\rangle\nonumber\\
&&+\left\langle \mathcal{FM}\left[\begin{array}{c}\hat{\psi}\\
\hat{\phi}(s)
\end{array}\right],\left[\begin{array}{c}\hat{\psi}\\
\hat{\phi}(s)
\end{array}\right]\right\rangle\nonumber\\
&&+\left\langle \left[\begin{array}{c}\hat{\psi}\\
\hat{\phi}(s)
\end{array}\right],\mathcal{FM}\left[\begin{array}{c}\hat{\psi}\\
\hat{\phi}(s)
\end{array}\right]\right\rangle.
\end{array}
\end{equation*}
To show that $\dot V<0$, we examine $\mathcal{A}\mathcal{P}$ and $\mathcal{FM}$ separately. First, we have
\begin{equation*}\label{22}
\mathcal{A}\mathcal{P}\left[\begin{array}{c}\hat{\psi}\\
\hat{\phi}(s)\end{array}\right]=\left[\begin{array}{c}\Psi\\
\Phi(s)\end{array}\right],
\end{equation*}
where
\begin{equation*}
\begin{array}{rcl}
\Psi&=&AP\hat{\psi}+\int_{-r}^{0}AQ(s)\hat{\phi}(s)ds+Br Q^{T}(-r)\hat{\psi}\nonumber\\
&&+BS(-r)\hat{\phi}(-r)+\int_{-r}^{0}BR(-r,\theta)\hat{\phi}(\theta)d\theta,\nonumber\\
\Phi(s)&=&r\dot{Q}^{T}(s)\hat{\psi}+\dot{S}(s)\hat{\phi}(s)+S(s)\dot{\hat{\phi}}(s)\nonumber\\
&&+\int_{-r}^{0}\frac{d}{ds}R(s,\theta)\hat{\phi}(\theta)d\theta.
\end{array}
\end{equation*}
Then,
\begin{equation*}
\begin{array}{rcl}
&&\left\langle\left[\begin{array}{c}\hat{\psi}\\
\hat{\phi}(s)
\end{array}\right],\mathcal{A}\mathcal{P}\left[\begin{array}{c}\hat{\psi}\\
\hat{\phi}(s)
\end{array}\right]\right\rangle\nonumber\\
&=&\int_{-r}^{0}\hat{\psi}^{T}\Psi ds +\int_{-r}^{0}\hat{\phi}^{T}(s)\Phi(s)ds\nonumber\\
&=&r \hat{\psi}^{T}AP \hat{\psi} +r\int_{-r}^{0}\hat{\psi}^{T}A Q(s)\hat{\phi}(s)ds+r \hat{\psi}^{T}Br Q^{T}(-r)\hat{\psi}\nonumber\\
&&+r \hat{\psi}^{T}B S(-r)\hat{\phi}(-r)+r\int_{-r}^{0}\hat{\psi}^{T}BR(-r,\theta)\hat{\phi}(\theta)d\theta\nonumber\\
&&+\int_{-r}^{0}r\hat{\phi}^{T}(s)\dot{Q}^{T}(s)\hat{\psi} ds+\int_{-r}^{0}\hat{\phi}^{T}(s)\dot{S}(s)\hat{\phi}(s)ds\nonumber\\
&&+\int_{-r}^{0}\int_{-r}^{0}\hat{\phi}^{T}(s)\frac{d}{ds}R(s,\theta)\hat{\phi}(\theta)dsd\theta \nonumber\\
&&+\int_{-r}^{0}\hat{\phi}^{T}(s)S(s)\dot{\hat{\phi}}(s)ds\nonumber\\
&=&\int_{-r}^{0}\left[\begin{array}{c}\hat{\psi}\\
\hat{\phi}(-r)\\
\hat{\phi}(s)\end{array}\right]^{T}\Sigma\left[\begin{array}{c}\hat{\psi} \\
\hat{\phi}(-r)\\
\hat{\phi}(s)\end{array}\right]ds\nonumber\\
&&+\int_{-r}^{0}\int_{-r}^{0}\hat{\phi}^{T}(s)\frac{d}{ds}R(s,\theta)\hat{\phi}(\theta)dsd\theta\nonumber\\
&&+\int_{-r}^{0}\hat{\phi}^{T}(s)S(s)\dot{\hat{\phi}}(s)ds,
\end{array}
\end{equation*}
where
\begin{equation*}
\Sigma=\left[\begin{array}{ccc}AP+r BQ^{T}(-r)&BS(-r)&\Theta\\
0 & 0 &0\\
r\dot{Q}^{T}(s)& 0 & \dot{S}(s)\end{array}\right],
\end{equation*}
where $\Theta=r (AQ(s)+BR(-r,s))$.

Since $\left[\begin{array}{c}\hat{\psi}\\
\hat{\phi}(s)\end{array}\right] \in X$, we have $\hat{\phi}(0)=C\hat{\psi}+D\hat{\phi}(-r)$. Then,
\begin{equation*}
\begin{array}{rcl}
&&\int_{-r}^{0}\hat{\phi}^{T}(s)S(s)\dot{\hat{\phi}}(s)ds\nonumber\\
&=&\hat{\phi}^{T}(0)S(0)\hat{\phi}(0)-\hat{\phi}^{T}(-r)S(-r)\hat{\phi}(-r)\nonumber\\
&&-\int_{-r}^{0}\hat{\phi}^{T}(s)\dot{S}(s)\hat{\phi}(s)ds-\int_{-r}^{0}\dot{\hat{\phi}}^{T}(s)S(s)\hat{\phi}(s)ds\nonumber\\
&=&\frac{1}{2}\left(\hat{\phi}^{T}(0)S(0)\hat{\phi}(0)-\hat{\phi}^{T}(-r)S(-r)\hat{\phi}(-r)\right)\nonumber\\
&&-\frac{1}{2}\int_{-r}^{0}\hat{\phi}^{T}(s)\dot{S}(s)\hat{\phi}(s)ds\nonumber\\
&=&\frac{1}{2}\int_{-r}^{0}\left[\begin{array}{c}\hat{\psi}\\
\hat{\phi}(-r)\\
\hat{\phi}(s)\end{array}\right]^{T}\Omega\left[\begin{array}{c}\hat{\psi}\\
\hat{\phi}(-r)\\
\hat{\phi}(s)\end{array}\right]ds.
\end{array}
\end{equation*}
where
\begin{equation*}
\Omega=\left[\begin{array}{ccc}\frac{1}{r}C^{T}S(0)C& \frac{1}{r}(C^{T}S(0)D)&0\\
\frac{1}{r}(D^{T}S(0)C)& -\frac{1}{r}(S(-r)-D^{T}S(0)D)& 0\\
0& 0& -\dot{S}(s)
\end{array}\right].
\end{equation*}
Thus,
\begin{equation*}
\begin{array}{rcl}
\dot{V}&=&\left\langle \left[\begin{array}{c}\hat{\psi}\\
\hat{\phi}(s)
\end{array}\right],\mathcal{A}\mathcal{P}\left[\begin{array}{c}\hat{\psi}\\
\hat{\phi}(s)
\end{array}\right]\right\rangle\nonumber\\
&&+\left\langle\mathcal{A}\mathcal{P}\left[\begin{array}{c}\hat{\psi}\\
\hat{\phi}(s)
\end{array}\right],\left[\begin{array}{c}\hat{\psi}\\
\hat{\phi}(s)
\end{array}\right]\right\rangle\nonumber\\
&&+\left\langle \mathcal{FM}\left[\begin{array}{c}\hat{\psi}\\
\hat{\phi}(s)
\end{array}\right],\left[\begin{array}{c}\hat{\psi}\\
\hat{\phi}(s)
\end{array}\right]\right\rangle\nonumber\\
&&+\left\langle \left[\begin{array}{c}\hat{\psi}\\
\hat{\phi}(s)
\end{array}\right],\mathcal{FM}\left[\begin{array}{c}\hat{\psi}\\
\hat{\phi}(s)
\end{array}\right]\right\rangle\nonumber\\
&=&\int_{-r}^{0}\left[\begin{array}{c}\hat{\psi}\\
\hat{\phi}(-r)\\
\hat{\phi}(s)\end{array}\right]^{T}\Delta\left[\begin{array}{c}\hat{\psi}\\
\hat{\phi}(-r)\\
\hat{\phi}(s)\end{array}\right]ds\nonumber\\
&&+\int_{-r}^{0}\int_{-r}^{0}\hat{\phi}^{T}(s)\left(\frac{d}{ds}R(s,\theta)\right.\nonumber\\
&&\left.+\frac{d}{d\theta}R(s,\theta)\right)\hat{\phi}(\theta)dsd\theta,
\end{array}
\end{equation*}
where
\begin{equation*}
\Delta=\left[\begin{array}{ccc}
\Gamma & B S(-r)+FM_{1}+\frac{1}{r}(C^{T}S(0)D) &\Upsilon\\
\ast&-\frac{1}{r}(S(-r)-D^{T}S(0)D) & 0\\
\ast & \ast &\dot{S}(s)
\end{array}\right].
\end{equation*}

From conditions (\ref{4})-(\ref{5}), we have
\begin{equation*}
\dot{V}<0.
\end{equation*}
Therefore, the closed-loop System (\ref{a1})-(\ref{a2}) is stable.
\end{pf}
\begin{rem}
When $F=0$ in System (\ref{a1})-(\ref{a2}), we recover the standard delay-differential framework studied in \cite{Peet:16} and \cite{Peet:13}:
\begin{eqnarray*}
\dot{x}(t)&=&A_{0}x(t)+\sum_{i=1}^{l}A_{i}x(t-r_{i})\nonumber\\
x(t)&=&\phi(t).
\end{eqnarray*}
The primary computational advantage of the differential-difference framework over control of System (\ref{a1})-(\ref{a2}) is that we can replace $A_i \in \R^{n \times n}$  with $BC$ where $B\in \R^{n \times m}$ and $C\in \R^{m \times n}$ and $m$ is typically strictly less than $n$. Because the dimension of the decision variables in the optimization problem defined in this paper scale as $n+2m$ as opposed to $3n$ using the framework in \cite{Peet:16} and \cite{Peet:13}, the complexity of the resulting algorithm is significantly reduced.
\end{rem}
\begin{rem}
The feedback controller in (\ref{13}) does not include delay in the input. However, the case of delay in the input can also be treated using a different form of $\mathcal{F}$. This is left for future work.
\end{rem}
\begin{rem}
Although not explicitly stated, in order to use SOS to enforce the conditions of Theorem 3 and Proposition 1, we choose our decision variables to be polynomial and use SOSTOOLS and the Positivstellensatz to enforce positivity/negativity on the interval $[-r,0]$. This approach is described in more detail in \cite{Peet:16} and \cite{Peet:13}.
\end{rem}

In the following, we present a numerical example to illustrate the controller obtained from the condition in Proposition 1.
We consider the following system with a feedback controller as follows
\begin{eqnarray}
\dot{x}(t)&=&\left[\begin{array}{cccccc} 0& 0.5 &0& 0& 0& 0\\
-0.5 &-0.5&0 & 0&0 &0\\
0& 1 &0.1 & 1 & 0& 0\\
0& 0& -2& 0.2 & 0 & 0\\
0 & 0 & 0 & 1 & -2 & 0\\
0 & 0 & 0 & 0 & 0 & -0.9
\end{array}
\right]x(t)\nonumber\\
&&+\left[\begin{array}{cc} 0.5 & 0\\
0 & 0\\
0 & 0\\
0 & 0\\
0 & 0\\
0 & 1
\end{array}
\right]y(t-r)+\left[\begin{array}{c}1\\
0\\
0\\
0\\
0\\
1
\end{array}
\right] u(t),\label{s1}\\
y(t)&=&\left[\begin{array}{cccccc} -0.2 & 0 & 0& 0 & 0 &0\\
0 & 0& 0& 0 & 0 & 1
\end{array}
\right]x(t),\label{s2}
\end{eqnarray}
where $r=1.6$s. By using Proposition 1, together with the tools of MuPad, Matlab, SOSTOOLS and polynomials with degree 2, we obtain the controller
\begin{eqnarray}\label{19}
u(t)&=&\left[\begin{array}{c}-1.874\\2.232\\-0.830\\3.099\\0.030\\-1.033
\end{array}\right]^{T}x(t)+\left[\begin{array}{c} -0.239 \\ -0.343\end{array}\right]^{T}y(t-r)\nonumber\\
&&+\int_{-1.6}^{0}K_{2}y(t+s)ds,
\end{eqnarray}
where
\begin{equation*}
K_{2}=\left[\begin{array}{c} -0.246+0.221s+0.122s^{2}-0.012s^{3}-0.032s^{4}\\
0.238-0.398s++0.007s^{2}+0.037s^{3}+0.010s^{4},
\end{array}
\right]^{T}.
\end{equation*}
Using Controller (\ref{19}) coupled with System (\ref{s1})-(\ref{s2}) we simulate the closed-loop system, which is illustrated in Fig.1.
\begin{figure}[thpb]
      \centering
\includegraphics[scale=0.6]{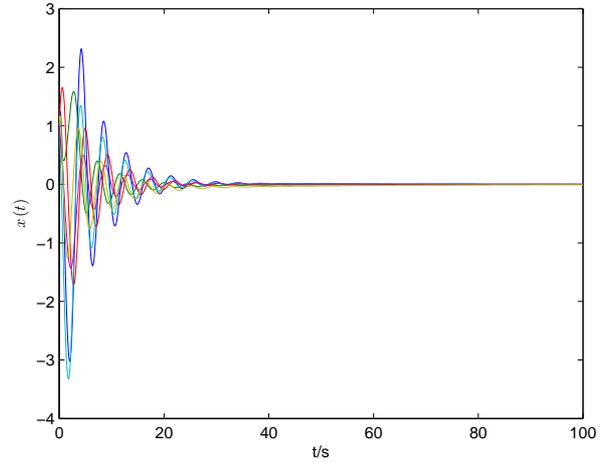}
      \caption{States of System (\ref{s1})-(\ref{s2}) coupled with stabilizing Controller from Prop. 1}
      \label{figurelabel}
\end{figure}
\section{CONCLUSIONS}
In this paper, we have obtained an analytic formulation for the inverse of jointly positive multiplier and integral operators as defined in \cite{Peet:16}. This formulation has the advantage that it eliminates the need for either individual positivity of the multiplier and integral operators or the need to use a series expansion to find the inverse. This inversion formula is applied to controller synthesis of coupled differential-difference equations. The use of the differential-difference formulation has the advantage that the size of the resulting decision variables is reduced, thereby allowing for control of systems with larger numbers of states. These methods are illustrated by designing a stabilizing controller for a system with 6 states and 2 delay channels..


\end{document}